\documentclass[twocolumn,showpacs,preprintnumbers,amsmath,amssymb,prl]{revtex4}

\usepackage[dvips]{graphicx}
\usepackage{dcolumn}
\usepackage{bm}
\usepackage{color}
\usepackage{amsmath}

\begin{document}

\preprint{[Phys. Rev. Lett. {\bf 57}, 1847 (1986)]}
\title{Repolarization of Negative Muons by Polarized $^{209}$Bi Nuclei}

\author{R. Kadono}
\affiliation{Department of Physics and Meson Science Laboratory, Faculty of Science, University of Tokyo, Tokyo 113, Japan}
\author{J. Imazato}
\affiliation{Department of Physics and Meson Science Laboratory, Faculty of Science, University of Tokyo, Tokyo 113, Japan}
\author{T. Ishikawa}
\affiliation{Department of Physics and Meson Science Laboratory, Faculty of Science, University of Tokyo, Tokyo 113, Japan}
\author{K. Nishiyama}
\affiliation{Department of Physics and Meson Science Laboratory, Faculty of Science, University of Tokyo, Tokyo 113, Japan}
\author{K. Nagamine}
\affiliation{Department of Physics and Meson Science Laboratory, Faculty of Science, University of Tokyo, Tokyo 113, Japan} 
\author{T. Yamazaki}
\affiliation{Department of Physics and Meson Science Laboratory, Faculty of Science, University of Tokyo, Tokyo 113, Japan}
\author{A. Bosshard}
\affiliation{Physik-Institut der Universit\"at Z\"urich, CH-8001 Z\"urich, Switzerland}
\author{M. D\"obeli}
\affiliation{Physik-Institut der Universit\"at Z\"urich, CH-8001 Z\"urich, Switzerland}
\author{L. van Elmbt}
\affiliation{Physik-Institut der Universit\"at Z\"urich, CH-8001 Z\"urich, Switzerland}
\author{M. Schaad}
\affiliation{Physik-Institut der Universit\"at Z\"urich, CH-8001 Z\"urich, Switzerland}
\author{P. Tru\"ol}
\affiliation{Physik-Institut der Universit\"at Z\"urich, CH-8001 Z\"urich, Switzerland}
\author{A. Bay}
\affiliation{Institut de Physique Nucl\'eaire, Universit\'e de Lausanne, CH-1015 Lausanne, Switzerland}
\author{J. P. Perroud}
\affiliation{Institut de Physique Nucl\'eaire, Universit\'e de Lausanne, CH-1015 Lausanne, Switzerland}
\author{J. Deutsch}
\affiliation{Institut de Physique Corpusculaire, Universit\'e Catholique de Louvain, B-1348 Louvain-la-Neuve, Belgium}
\author{B. Tasiaux}
\affiliation{Institut de Physique Corpusculaire, Universit\'e Catholique de Louvain, B-1348 Louvain-la-Neuve, Belgium}
\author{E. Hagn}
\affiliation{Physik Department, Technische Universit\"at M\"unchen, D-8046 Garching, Federal Republic of Germany}
\date{Received 28 April 1986, Corrected 7 October 2016}
\begin{abstract}
A large $\mu^-$ polarization was achieved in muonic Bi atoms with the help of the strong hyperfine field in a polarized nuclear target.  Using $^{209}$Bi nuclei polarized to ($59\pm9$)\%  in ferromagnetic BiMn, we observed a $\mu$-$e$ decay asymmetry of ($13.1\pm3.9$)\%,  which gives $\mu^-$  polarization per nuclear polarization equal to $-1.07\pm 0.35$.  This value is almost consistent with $-0.792$ calculated for nuclei with spin $I= \frac{9}{2}$ and a positive magnetic moment  under the assumption  that the hyperfine interaction becomes effective in the lowest muonic states.
\end{abstract}
\maketitle

Since the development of the intense proton-accelerator facility known as the meson factory, positive and  negative muons  ($\mu^+,\mu^-$) have been two of the most  fruitful   probes  in  various  fields  of  physics.\cite{Hughes:77}
However, because of the small spin polarization of the negative muon  ($\mu^-$) after the formation  of a muonic atom  (which is always the case for $\mu^-$ in matter),  experimental investigation by the $\mu^-$ is still in the dawn in contrast to the established value of the $\mu^+$ in fundamental  physics, condensed  matter  physics, and so on. Actually, the residual polarization of $\mu^-$ ($= P_\mu$) in the $1s$ ground state of a muonic atom is reduced to about $\frac{1}{6}$of its initial value through the atomic capture into a
fine-structure doublet and the succeeding cascade process  down  to  the  $1s$ state.\cite{Mann:61}  In  the  case  of  muonic atoms of nonzero-spin nuclei, there exists an additional loss of the polarization due to the hyperfine (hf) interaction, which reduces the final  $P_\mu$ to less than 5\%.\cite{Uberall:59}
It is absolutely  clear  that  the  introduction  of a  new method  to increase  $P_\mu$  for  the $\mu^-$ will enable us to develop many interesting studies using the $\mu^-$.

In this paper we shall report an experimental  study to realize the high spin polarization of negative muons. The principal idea of artificial polarization (i.e., ``repolarization") for the $\mu^-$ was proposed ten years ago\cite{Nagamine:76}  as a  consequence   of   the  study   of   polarized  muonic atoms.\cite{Nagamine:74}    The  process of repolarization  consists of  (a) the atomic capture of $\mu^-$  by a polarized nucleus forming a muonic atom,  (b) the hf coupling of the $\mu^-$  spin with the polarized nuclear spin in some muonic states, {c) the atomic transition  by $E1$  cascades down to the hf doublet of the $1s$ state  $F^\pm$ ($= I\pm\frac{1}{2}$)  where $I$ refers to the  nuclear spin),  and {d) the final  $M1$  transition from the higher hf state to the lower one.  On the assumption  that  the  initial $\mu^-$   polarization  is equal  to zero and  that the hf coupling is active only at the $1s$ state,  the residual $\mu^-$   polarization in pure  $F^\pm$  states after step (c) is calculated to be
\begin{eqnarray}
P_\mu(F^+) &=& [(2I+3)I/(I + 1)(2I+ 1) ]P_N, \\\nonumber
P_\mu(F^-) &=& -[(2I-1)/(2I+1)]P_N,\nonumber
\end{eqnarray}
where $P_N$ denotes the nuclear polarization.  In the case of light nuclei  (atomic  number $Z$ below 10),  the last step  (d)  is slow compared  with the $\mu^-$ lifetime\cite{Wilson:61}  and the population-averaged sum of $P_\mu(F^+)$ and $P_\mu(F^-)$ (i.e., $P_\mu$) remains small because of the opposite $\mu^-$polarizations in the respective states.\cite{Yamazaki:80,Hambro:77}  On the other hand, step (d) becomes much faster than the muon disappearance rate in heavier nuclei ( $\sim$10$^{-9}$ sec) and $P_\mu$ approaches
$$P_\mu(F^-)=- [4I(I + 1)(2I-1)/(2I +1)^3]P_N$$
for positive $\mu_I$,
$$P_\mu(F^+) = [4I^2 (2I+ 3)/(2I+ 1)^3]P_N$$
for negative $\mu_I$,  where  $\mu_I$ refers to the magnetic moment  of  the  nucleus.\cite{Nagamine:76,Nagamine:74,Yamazaki:80}   Note  that  the  effect increases  with the nuclear  spin $I$; for instance, the above formula  gives $P_\mu= -0.792P_N$  for  $I=\frac{9}{2}$ with a positive nuclear  moment.

It should  be noted  that the observation of such  a repolarization  effect  itself  is important enough to get  a more  correct  understanding of  the  fundamental  properties  of muonic  atoms.   There are some  indications in light  nuclei  that  the  hf coupling  is ``switched on" in muonic  orbits  higher  than  the  $1s$ one.\cite{Favart:70}    Recently, the influence of  this  effect  on  $P_\mu$  was also  calculated.\cite{Kuno:86} The  calculation  shows  that the residual  $P_\mu$ depends on the   assumed  state   where   the  hf splitting   becomes larger than the decay width.

As already suggested  in the original  proposal,\cite{Nagamine:76} a ferromagnetic compound BiMn ($T_c = 633$ K)  was chosen for the agent of the polarized $^{209}$Bi ($I= \frac{9}{2}$) nuclear  target.    The    strong    internal    magnetic    field   $H_{\rm int}$  of $+940^{+180}_{-130}$ kG at Bi sites\cite{Nagamine:72,Koyama:77} and   $-224.5$ kG  at  Mn sites\cite{Hihara:70}  is suitable  to realize  the  large  nuclear  polarization  by the  thermal  equilibrium method.  The  characteristic   temperature  $T_0 =\mu_IH_{\rm int}/ k_BI$ for  the   nuclear polarization   is $T_0 = 31(5)$ mK  for  Bi and  11  mK  for Mn,  both  of which are  readily achievable  with a $^3$He-$^4$He dilution refrigerator.  For a sample  with a thin disk shape,  an  external field  of 5--8 kG  is required  to get saturation  magnetization along  the  disk  plane.   The target  polarization  was monitored by two carbon  resistance   thermometers   which   were    calibrated in   a separate   experiment  by  $\gamma$-ray  anisotropy    measurements of  $^{207}$BiMn  dispersed   in  a  small  disk  of  the BiMn.\cite{Koyama:77}

The  goal of this repolarization experiment was to determine the  degree  of $\mu^-$ polarization  acquired   by unpolarized muons  in muonic  $^{209}$Bi atoms,  by measurement  of   the   decay-electron  asymmetry  for   a known  target  polarization.  The  experiment was  performed  at the $\mu E1$ channel  of the Swiss Institute for Nuclear  Research  (SIN).   The configuration of the  $e^-$ detector system  is shown  schematically  in Fig. 1.  The magnetic  field was applied  in the direction  perpendicular to the $\mu^-$ beam,  eliminating any effect  on  the  $e^-$ asymmetry coming  from  the  initial  polarization  of the $\mu^-$ beam.   Two  Nal  detectors were  installed  to  measure  the energy  spectrum of the decay $e^-$ with a total solid angle of about 1\% for each side.  The  trigger condition   for   the  ``good" $\mu$-$e$  event  was  $(\mu1\cdot \mu2\cdot\mu3\cdot \overline{\mu4})\cdot$[(TR1$\cdot$TR2$\cdot$NaI-R) or (TL1$\cdot$TL2$\cdot$NaI-L)].

Negative  muons  of 115  MeV/$c$ (partly  85  MeV/$c$) were  stopped   with  a rate  of  $2\times10^5$/sec in  the  target, whereas  the    typical    trigger    rate    was   about    60 events/sec.  In  order  to  improve  the  signal-to-background  ratio, four  planes of multiwire  proportional chambers were used  to trace  back the  $e^-$  trajectory. A Ge(Li) detector  was placed near  the  target to monitor  the  number of the stopped  muons  and  the atomic capture  ratio of $\mu^-$ in and around  the target.
 
To avoid systematic  errors  in our  measurement, the degree  of target  polarization  was varied  by changing  of the  target  temperature without  any  other  changes   in the experimental conditions. We repeated the measurements four  times,  each cycle consisting  of pairs of experiments performed at both $62\pm4$ mK and 1.0--4.2 K.  The  low-temperature experiments correspond to a polarization  of ($59\pm 9$)\%  for  Bi and  ($- 21 \pm1$)\%  for Mn.  It took about  24 h to complete a pair of the measurements, during which the stability of the detector system  was confirmed by the  constancy  of the  trigger ratio  between  the  $R$ and  the $L$ sides  (see  Fig.~1).   Additional  split  beam counters allowed  us to monitor  any left-right  shift  of  the  beam;  none  was observed.  One of  the  four  cycles  was done  with  the  magnetic  field direction reversed.
\begin{figure}[t]
	\begin{center}
\includegraphics[width=0.95\linewidth]{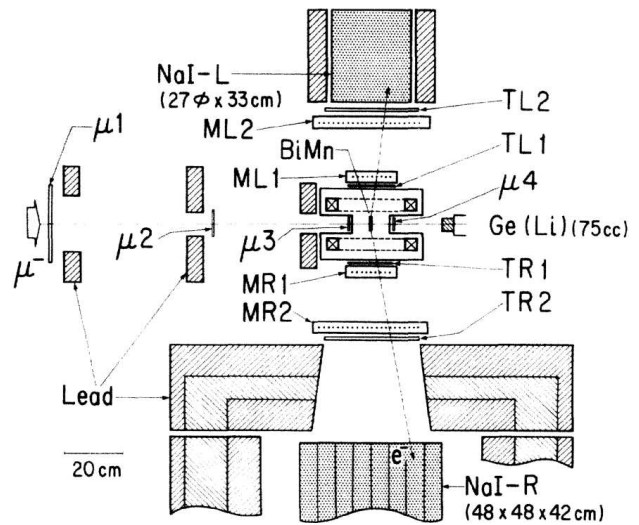}
			\caption{Top view of the experimental  setup.  The letters $\mu$, T, and M refer to muon counters,  electron counters,  and multiwire   proportional  chambers,   respectively.   R  and  L denote  right and left of the target as seen coming with the muon  beam.  Nal indicates the sodium  iodide and Ge(Li)  a high-resolution semiconductor detector.}
			\label{fig1}
	\end{center}
\end{figure}

The   observables  which   were   used   for   the   final analysis are as follows:   (a) the time spectra  of electron decay  events with  respect   to  the $\mu^-$ stop   (up  to  2 $\mu$sec), (b)  the time spectra  of the decay event  relative to the $\mu1$ counter alone  (up to 200 nsec),  (c) the time spectra  of  the  muon  pileup  signal  during  2 J.tSec, (d) the  electron energy  spectra  of  the  Nal  detectors, and
(e)   the   hit   patterns  of   the   multiwire  proportional chambers.  The  data acquisition was done  with a PDP-11 computer and the information was recorded  on magnetic  tapes on an event-by-event mode.
In the off-line  analyses,  the efforts were directed  toward the improvement of the signal-to-background  ratio  in  the  time  spectra  (a).   All the  events which  did not  hit  the  multiwire   proportional chambers  properly were removed.   The  background in the ``prompt" region of the spectra (a), probably coming from the contaminant  $e^-$ in the  beam, was reduced  by use of the information from  (b).  The effect of accidental secondary muons distorting  the time spectra was eliminated by the rejection of events recorded in the spectra (c).

Because of the substantial  energy loss in the target (an average of 15 g/cm$^2$ in the direction of the electrons)  as well as the distortion of the decay spectrum from  the  bound  muons,\cite{Gilinsky:60,Huff:61,Haenggi:74}  the  detected  electrons have an energy spectrum concentrated in the lower energy region.  The decay events which have energy higher than 35 MeV have to originate from low-$Z$ contaminants  (Al walls, carbon in the detectors) and were removed  to improve the signal-to-background ratio in the spectra  (a).  By use of a maximum-likelihood method,\cite{Eadie:71}  the spectra of (a) after  these cuts were fitted with the function
$$N(t) = \sum_i N^i \exp(- t/\tau_i) + B,$$
where  $N^i$  refers  to  the counting  rate of electrons  at $t = 0$ for the $i$th component  which has the lifetime $\tau_i$, and $B$ refers to the constant background.  The assumed components  are Bi ($\tau = 73$ nsec),  Mn ($\tau= 225$ nsec), Al ($\tau = 865$ nsec), and C ($\tau= 2038$ nsec),  where the lifetimes  were  fixed  and  the  $N^i$'s  were  determined through the fitting.  A typical example of the spectrum is shown with a result of fitting in Fig. 2.

The  time spectra of  both  $R$ and $L$ sides were analyzed independently and yield for each cycle of experiments   the   amplitudes   $N^i_R$ (62 mK),   $N^i_L$ (62  mK), $N^i_R$ (4.2 K),  and   $N^i_L$ (4.2 K).   We  shall  discuss  the analysis of  the  Bi components  which have  the  most precisely determined amplitudes.

\begin{figure}[b]
	\begin{center}
\includegraphics[width=0.95\linewidth]{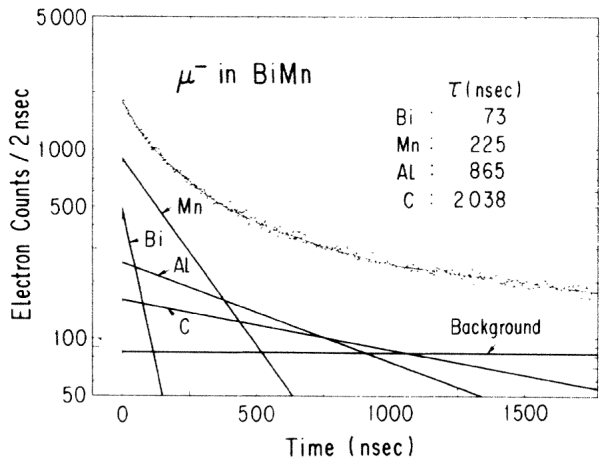}
			\caption{A typical $\mu$-$e$ decay-time spectrum obtained after the various background cuts discussed in the text. The time resolution was 6--10 nsec. The muons are assumed to be captured by Bi, Mn, Al, and C, where the last two components are ascribed to muons stopped in the cryostat walls and in the scintillators.}
			\label{fig2}
	\end{center}
\end{figure}

\begin{table}[h]
\begin{tabular}{ccccc}
\hline\hline
Cycle & $D$ & $C_R$ & $C_L$ & $A_{\rm exp}$ \\
\hline
$1(+)$ & $1.147(131)$ & ${\color{red} +}1.011(19)$ & ${\color{red} -}0.419(11)$ & ${\color{red} +}0.098(84)$ \\
$2(+)$ & $1.191(85)$ & ${\color{red} +}0.644(16)$ & ${\color{red} -}0.430(13)$ & ${\color{red} +}0.165(70)$ \\
$3(-)$ & $0.871(78)$ & ${\color{red} -}0.681(17)$ & ${\color{red} +}0.497(12)$ & ${\color{red} +}0.116(74)$ \\
$4(+)$ & $1.175(107)$ & ${\color{red} +}0.692(15)$ & ${\color{red} -}0.443(13)$ & ${\color{red} +}0.144(83)$ \\
 & & & & \\
 & & & Average & ${\color{red} +}0.131(39)$\\
\hline\hline
\end{tabular}
\caption{The numerical results of the analysis for the Bi
component. The sign behind the cycle number refers to the
polarity of the polarizing magnetic field. For abbreviations,
see text. The quoted error of $C_{R,L}$ comes from statistics of
the Monte Carlo calculation.}
\end{table}

If the polarization acquired by the muons is $P_\mu$, the emitted  electrons are distributed  as a function of their energy  $E$ and  emission  angle  $\theta$   relative  to  $P_\mu$  as
$\alpha(E)+\beta(E)P_\mu\cos\theta$.   For  the  functions  $\alpha(E)$  and $\beta(E)$  we assumed  their  theoretical  forms  extracted from  Refs.\cite{Gilinsky:60,Huff:61,Haenggi:74};  let us note  that we obtain for the total  asymmetry,   integrated    between    $E = 0$   and $E= E_{\rm max}$,
$$A_0= \int_0^{E_{\rm max}}\beta(E)dE\left[\int_0^{E_{\rm max}}\alpha(E)dE\right]^{-1} = -0.21.$$
We compute now with Monte Carlo methods the acceptance of our detectors for both the isotropic and the
nonisotropic  distribution   functions  $\alpha(E)$ and $\beta(E)$, taking into account also the energy cuts used in each cycle to construct  the time distribution  (a).  Then,  we can express  $A_{\rm exp} = P_\mu A_0$ as a function  of the experimentally measured amplitudes as
$$A_{\rm exp} =(1-D)/(DC_L-C_R),$$
where $D$ is the double ratio [$N_R(62$ mK)/$N_R(4.2$ K)][$N_L(62$ mK)/$N_L(4.2$ K)]$^{-1}$ and $C_{R,L}$ is the correction factor from the Monte Carlo calculation.
\begin{table}[b]
\begin{tabular}{ccc}
\hline\hline
$hf$ switched on & \hspace{3cm} & $P_\mu/P_{\rm Bi}$ \\
\hline
At $1s$ state & &$-0.792$ \\
At $2p$ state & &$-0.750$ \\
At $3d$ state & &$-0.688$ \\
 & &  \\
Expt. & & $-1.07\pm0.35$\\
\hline\hline
\end{tabular}
\caption{The  repolarization  efficiency  of  muons  in muonic  209Bi atoms  as  a  function  of  muonic  states  from where hyperfine coupling becomes active.  The initial polarization of the muon is assumed to be zero.}
\end{table}

In Table I the summary of the analysis is shown for Bi results.  A preliminary result was reported elsewhere.\cite{Kadono:86} From  $A_{\rm exp}= -0.21P_\mu = {\color{red} +}0.131\pm0.039$  we obtain  the  repolarization  coefficient   $P_\mu/ P_{\rm Bi} = - 1.07\pm0.35$  which proves the feasibility of the method suggested in Refs.\cite{Nagamine:76} and \cite{Nagamine:74}.

Our result is compared in Table II with the theoretical values obtained with the assumption of the onset of the hyperfine interaction in various muonic-mesic levels.  Because of the moderate  precision we cannot distinguish between these scenarios; it would be also premature  to do so before the theoretical assumption $A_0 = -0.21$ is experimentally assessed (such an experiment is under way by our collaboration at SIN).

In summary,  the present experiment  has demonstrated for the first time that negative muon spin can be repolarized by the aid of nuclear polarization.  This method    will   be   powerful   in   studying   $\mu^-$-spin-dependent  phenomena.

We  thank  Dr.  S.  Arai  for  the  preparation  of  the BiMn samples, Dr. R. S. Hayano for his help compressing data, and Dr. Y. Kuno for helpful discussion.  Also we wish to thank Prof. J. P. Blaser for his hospitality at SIN, and  the staff of SIN for their  help with our experiment.   The experiment  was partly supported by a Grant-in-Aid  of the Japanese Ministry of Education, Culture and Science.

\end{document}